\documentclass[preprint]{elsarticle}

\usepackage{hyperref}
\usepackage{amsmath}
\usepackage{amsfonts}
\usepackage{amssymb}
\usepackage{siunitx}
\usepackage{multirow}
\usepackage{graphicx}
\usepackage{soul}
\usepackage{todonotes}

\hypersetup{pdfauthor={Name}}

\begin{document}

\begin{frontmatter}

\title{Radial Oscillations and Dynamical Instability Analysis for Linear-Quadratic GUP-modified White Dwarfs}

\author[add1,add3]{John Paul R. Bernaldez\corref{cor1}}
\ead{bernaldezjp@gmail.com}
\author[add2]{Adrian G. Abac}
\ead{adrian.abac@aei.mpg.de}
\author[add1]{Roland Emerito S. Otadoy}
\ead{rsotadoy@usc.edu.ph}

\cortext[cor1]{Corresponding author}
\address[add1]{Theoretical and Computational Sciences and Engineering Group, Department of Physics, University of San Carlos, Cebu City, Philippines 6000}
\address[add3]{Faculty of Physics and Astronomy Friedrich-Schiller-Universit{\"a}t Jena, 07743 Jena, Germany}
\address[add2]{Max Planck Institute for Gravitational Physics (Albert Einstein Institute), Am M{\"u}hlenberg 1, D-14476 Potsdam, Germany}


\begin{abstract}
	A modification to the Heisenberg uncertainty principle is called the generalized uncertainty principle (GUP), which emerged due to the introduction of a minimum measurable length, common among phenomenological approaches to quantum gravity. One approach to GUP is called linear-quadratic GUP (LQGUP) which satisfies both the minimum measurable length and the maximum measurable momentum, resulting to an infinitesimal phase space volume proportional to the first-order momentum $(1 - \alpha p)^{-4} d^3x d^3p$, where $\alpha$ is the still-unestablished GUP parameter. In this study, we explore the mass-radius relations of white dwarfs whose equation of state has been modified by LQGUP, and provide them with radial perturbations to investigate the dynamical instability arising from the oscillations. We find from the mass-radius relations that the main effect of LQGUP is to worsen the gravitational collapse by decreasing the mass of the relatively massive white dwarfs (including their limiting mass, while increasing their limiting radius). This effect gets more prominent with larger values of $\alpha$. We then compare the results with available observational data. To further investigate the impact of the GUP parameter, a dynamical instability analysis of the white dwarf was conducted, and we find that instability sets in for all values of $\alpha$. With increasing $\alpha$, we also find that the central density at which instability occurs decreases, resulting to a lower maximum mass. This is in contrast to quadratic GUP, where instability only sets in below a critical value of the quadratic GUP parameter.
\end{abstract}

\begin{keyword}
	white dwarfs; GUP; mass-radius relation; equation of state.
	
	
\end{keyword}

\end{frontmatter}



\section{Introduction}\label{section: Introduction}
	
	A modification to the fundamental commutation relation, which is predicted by several quantum gravity phenomenology approaches leads to the generalization of the Heisenberg's uncertainty principle (HUP). This modification to the HUP is called the generalized uncertainty principle (GUP), which emerged due to the introduction of a minimum measurable length \cite{Amati:1988tn,Blau:2009fzi,Gross:1987ar,Kempf:1994su,Scardigli:1999jh,Tawfik:2014zca,Tawfik:2015rva,Das:2008kaa} Several variants to GUP have been developed over the years, including the first of which was the so-called quadratic GUP (QGUP), which is consistent with string theory phenomenology \cite{Amati:1988tn,Blau:2009fzi,Gross:1987ar} and black hole physics \cite{Scardigli:1999jh}, and which leads to a phase space volume measure proportional to the square of momenta, i. e. $\left(1 + \beta p^2 \right)^{-3}d^3xd^3p$ \cite{Chang:2001bm}. More recently, a specific form to the generalized uncertainty principle, which satisfies not only black hole physics but also string theory and doubly special relativity (DSR), was developed and is called the ``linear-quadratic" GUP (LQGUP) approach \cite{Ali:2010yn,Ali:2011ap,Vagenas:2019wzd, Abac:2020drc}, which satisfies both the minimum measurable length and the maximum measurable momentum conditions. An invariant infinitesimal phase space volume proportional to first-order momentum, i.e. $\left(1-\alpha p\right)^{-4}d^3x d^3p$, can be derived from the LQGUP, where, in this case, the approach is known as the ``linear" GUP approach (indicated by the phase space volume correction only up to order $p$) \cite{Ali:2010yn,Vagenas:2019wzd,Abac:2020drc,Abac:2021jpf}. Here, $\alpha$ is the quantum gravity (or LQGUP) parameter, which has no established values, but with upper bounds ranging from $\alpha_0 < {10}^{12}-{10}^{24}$ where $\alpha_0 = \alpha \hbar/l_p$ is the dimensionless LQGUP parameter and $l_p$ is the Planck length. From this point forward in the study, however, for the sake of consistency, we refer to the GUP used as linear-quadratic GUP (LQGUP).
 
    LQGUP has been studied and applied to numerous systems, such as the internal structure of compact objects like white dwarfs \cite{Abac:2020drc, Moussa:2015yqy} and neutron stars \cite{Abac:2021jpf,Prasetyo:2022uaa}, where LQGUP modifies the equation of state (EoS), and correspondingly, the mass-radius relations. Several studies which involve investigating the effects of LQGUP on white dwarfs show that LQGUP resists collapse, by increasing or completely removing \cite{Ali:2013ii} their mass limit (Chandrasekhar mass\cite{Chandrasekhar:1931ih,Chandrasekhar:1934in,Chandrasekhar:1964zza}), while other studies show that LQGUP worsens the collapse by reducing the maximum mass limit and increasing the minimum radius limit \cite{Abac:2020drc,Moussa:2015yqy}. It should be noted that there are currently no established values for the GUP parameter $\alpha_0$. From Ref. \cite{Silbar:2003wm}, the upper bound of $\alpha_0$ is suggested to be in the order $10^{17}$ to be compatible with the electroweak theory. On the other hand, Ref. \cite{Kilic:2006as} suggests an upper bound for the GUP parameter of $\alpha_0 < 10^{24}$, in order to address the concerns with regards Landau energy shifts for particles that have a mass $m$ and a charge $e$ within a constant magnetic field and cyclotron frequency. The effects of linear-quadratic GUP on the Lamb shift also suggest for a GUP parameter upper bound of $\alpha_0 <10^{12}$ \cite{Das:2008kaa,Ali:2010yn,Ali:2011ap}. From Ref. \cite{Kepler:2006ns}, the gravitational wave event GW150914 suggests an upper bound of $\alpha_0 < 1.8 \times 10^{20}$. From Ref. \cite{Scardigli:2014qka,Vagenas:2017vsw}, the obtained bounds for the GUP parameter $\alpha_0$ from gravitational interaction is found to be in the range $\alpha_0<10^{10} \text{ and } \alpha_0<10^{35}$. In addition, from Ref. \cite{Amati:1988tn,Vagenas:2017vsw,Alsaleh:2017ttv,Scardigli:2016pjs}, an $\alpha_0$ value of the order unity is believed based on predictions from string theory. As pointed out in Ref. \cite{Bosso:2023aht} with their review on the constraints of the quadratic GUP parameter, it was discussed that the best bound for the dimensionless quadratic GUP parameter is $\beta_0 < 10^{33}$, which is from scanning tunneling microscope studies. Assuming the relationship $\beta_0 \sim \alpha_0^2$ implies an upper bound $\alpha_0 < 10^{17}$. However, as demonstrated by the other studies \cite{Abac:2020drc,Moussa_2015}, GUP effects only become significantly noticeable for $\alpha_0 > 10^{17}$, and values of $\alpha_0$ less than the bound will be dominated by numerical uncertainty when doing a full solution of the EoS and the white dwarf's structure equations, so that the values of $\alpha_0$ are chosen to better show the effects of LQGUP.
	In the context of general relativity, which matters for the most massive white dwarfs \cite{Schutz:2003nr}, and where GUP effects are known to be most prominent \cite{Abac:2020drc,Mathew:2017drw,  Abac:2021txj}, the dynamical instability of a white dwarf or a compact object can be investigated by studying its behavior under radial oscillations \cite{Chandrasekhar:1931ih}. This was done before in the context of the QGUP done by Ref. \cite{Mathew:2018nvo}, where, above some critical value of the dimensionless QGUP parameter $\beta_0$, QGUP resists gravitational collapse by increasing the Chandrasekhar mass limit \cite{Mathew:2018nvo,Wang:2010ct,Wang:2011iv,Rashidi:2015rro,Ong:2018zqn}. The GUP-modified EoS prevents further gravitational collapse and limits the formation of compact astrophysical objects to that of the white dwarfs, which is in contrast with astrophysical observations, since heavier compact objects such as neutron stars exist.  However, they find that dynamical instability can still set in for $\beta_0 \le 5.38 \times 10^{39}$ \cite{Mathew:2020wnx}. Since it was shown by Ref. \cite{Abac:2020drc,Abac:2021jpf,Moussa:2015yqy}, that LQGUP seems to worsen gravitational collapse, it is reasonable to explore dynamical instability on linear-quadratic-GUP-modified white dwarfs, and to see whether this effect is present for a large range of values of the LQGUP parameter. 
	
	We then present the main problem and objectives of this work --- investigating dynamical instability for LQGUP-modified white dwarfs. The study is structured as follows. We take the framework of general relativity (GR) and calculate the stellar structure of the LQGUP-modified white dwarfs, where, in Section \ref{section: Linear quadratic GUP-modified White Dwarfs} we will obtain the exact forms of the LQGUP-modified number density, energy density, and pressure of ideal Fermi gases at zero-temperature, and from there, a modified Chandrasekhar equation of state (EoS) will be formulated. We choose an ideal, (degenerate) electron gas since we are mainly interested in the phenomenological effects on LQGUP on white dwarfs. This assumption is sufficient for massive white dwarfs since the effects of the GUP modification manifests in the high-mass regime and the effects of finite temperature only occur at low densities and pressures \cite{Abac:2020drc,Boshkayev:2015sjk}. From the equation of state, in Section \ref{section:White Dwarf Structure} we obtain the analytical form of the LQGUP-modified structure equations for zero-temperature white dwarfs, and their solutions are found by using the Tolman-Oppenheimer-Volkoff (TOV) structure equations in general relativity (GR). From the structure equations, we obtain the modified mass-radius relation of the white dwarfs. We then carry out in Section \ref{section: Dynamical Instability Analysis} the dynamical instability analysis on the LQGUP-modified white dwarfs so that the maximal stable configuration of the system is identified. We compare the results obtained to the work of Ref. \cite{Mathew:2020wnx}, where QGUP-modified white dwarfs are found to have both unstable configurations and super-stable configurations (where instability does not set in, for various values of the GUP parameter), and white dwarfs can have arbitrarily large masses. Finally, we present in Section \ref{chapter: Conclusions and Recommendations} the conclusions and recommendations for future work.
	
	\section{Linear quadratic GUP-modified White Dwarfs}
	\label{section: Linear quadratic GUP-modified White Dwarfs}
	In this section, we review the Linear Generalized Uncertainty Principle (LQGUP), as well as the statistical mechanics of Fermi gases modified by LQGUP.
	\subsection{Generalized Uncertainty Principle}
	\label{section: Generalized Uncertainty Principle}
	The generalized uncertainty principle (GUP) is a modification to the Heisenberg's uncertainty principle, which emerged due to the existence of a minimum measurable length \cite{Amati:1988tn, Tawfik:2014zca}.  The fundamental commutation relation that satisfies both black hole physics and string theory is \cite{Kempf:1994su, Chang:2001bm, Mathew:2017drw}:
	\begin{equation}
	[x,p]=i \hbar (1+\beta p^2),
	\end{equation}
	which then results to the quadratic GUP given by:
	\begin{equation}
	\Delta x \Delta p \ge \frac{\hbar}{2} [1+\beta (\Delta p)^2+ \langle p \rangle^2].
	\end{equation}
	Another approach to the general uncertainty principle, which not only satisfies black hole physics but also fits well with string theory, and doubly special relativity (DSR), was developed and is called the “linear-quadratic” GUP (LQGUP) approach \cite{Ali:2010yn, Ali:2011ap, Vagenas:2019wzd}, where the commutation relation is given by:
	
	\begin{equation}
	\begin{split}
	\left[x_i,p_j\right]&=i\left[\delta_{ij}-\alpha\left(p\delta_{ij}+\frac{p_ip_j}{p}\right)+\alpha^2\left(p^2\delta_{ij}+3p_ip_j\right)\right]\\
	[x_i,x_j]&=[p_i,p_j]=0
	\end{split}
	\end{equation}
	
	and the generalized uncertainty principle in one dimension is given by:
	\begin{equation}
	\Delta x \Delta p \ge \frac{\hbar}{2} [1-2 \alpha \langle p \rangle +4 \alpha^2 \langle p \rangle^2],
	\end{equation}
	where $\alpha=\frac{\alpha_0}{\left(M_pc\right)}=\alpha_0\left(\frac{l_p}{\hbar}\right)$ is the LQGUP parameter \cite{Ali:2010yn,Ali:2011ap,Abac:2020drc,Cortes:2004qn}, $M_p$ is the Planck mass, and $l_p$ is the Planck length.

	\subsection{Statistical Mechanics of Fermionic Gases with LQGUP Modification}
	\label{section: Statistical Mechanics of Fermionic Gases with LQGUP Modification}
	Statistical mechanics plays an important role to understanding the structure of stars and compact objects since white dwarfs are generally composed of degenerate electron gases. With this, there is a need to derive the thermodynamic properties of the degenerate Fermi gas at zero temperature obeying Pauli’s exclusion principle. We start with the grand canonical ensemble \cite{Abac:2020drc,Moussa_2015,Mathew:2018nvo,Bertolami:2009wa} formalism in statistical mechanics in order to derive the relevant thermodynamic quantities, which are the number density $n$, energy density $e$, and pressure $P$. The partition function can be written as:
	\begin{equation}
	\ln Z = \sum_j \ln[1+e^{\frac{(\mu-E_j)}{k_B T}}].
	\end{equation}
	For large volumes, as modified by linear-quadratic GUP, we then have:
	\begin{equation}
	\int \frac{d^3 x d^3 p}{(2 \pi \hbar)^3} \rightarrow\int \frac{d^3 x d^3 p}{(2 \pi \hbar)^3(1-\alpha p)^4}. 
	\end{equation}
	Thus, the grand canonical potential can then be rewritten as:
	\begin{equation}
	\Phi = - g k_B T \int \frac{d^3 x d^3 p}{(2 \pi \hbar)^3 (1-\alpha p)^4} \ln [1+e^{\frac{(\mu-E)}{(k_BT)}}],
	\end{equation}
	where $g$ refers to the multiplicity of states due to the spin of the particle, and $(2 \pi \hbar)^3$ refers to the volume occupied by and energy state in phase-space. From this, we then derive the expressions for the number density $n$, pressure $P$, electron energy density $\epsilon_e$, and total energy density $\epsilon$. The expression for number density, pressure, and electron energy density for degenerate ideal Fermi gases are given by \cite{Abac:2020drc}:
	\begin{equation}\label{Eq8}
	n=\frac{8 \pi}{(2\pi \hbar)^3}\int_0^{p_F}\frac{p^2 dp}{(1-\alpha p)^4} 
	\end{equation}
	\begin{equation}\label{Eq9}
	\epsilon_e=\frac{8 \pi}{(2\pi \hbar)^3}\int_0^{p_F} E\frac{p^2 dp}{(1-\alpha p)^4}
	\end{equation}
	\begin{equation}\label{Eq10}
	P=\frac{8 \pi}{(2\pi \hbar)^3}\int_0^{p_F}(E_F-E)\frac{p^2 dp}{(1-\alpha p)^4} = E_Fn - \epsilon_e ,
	\end{equation}
	where $p_F$ is the Fermi momentum, which corresponds to the maximum possible momentum, corresponding to the Fermi energy $E_F$ in the zero-temperature case. To optimize the numerical computation, we define a set of dimensionless quantities:
	\begin{equation}\label{Eq11}
	\begin{split}
	\xi\equiv\frac{p_F}{m_ec}\\
	\widetilde{p}\equiv\frac{p}{m_ec}\\
	\sigma \equiv m_ec \alpha
	\end{split}
	\end{equation}
    Note that our chosen values of $\alpha \in [10^{17}, 10^{19}, 10^{20}, 7\times10^{20}]$ (for comparison with previous studies \cite{Abac:2020drc, Abac:2021jpf, Mathew:2017drw, Mathew:2018nvo, Mathew:2020wnx} and to make the GUP effects more noticeable) correspond to $\sigma \in [4.2\times 10^{-6}, 4.2\times 10^{-4}, 4.2\times 10^{-3}, 2.9\times 10^{-2}]$. Therefore, from this redefinition of the variables, we see that the LQGUP parameter becomes $\sigma < 1.0$ with the dimensionless momentum $\tilde{p}$ compensating for it by becoming relatively large (by a factor $m_e c$), allowing for more convenient and efficient numerical calculations. Such rescaling of the variables has also been done in Ref. \cite{Abac:2020drc, Abac:2021jpf, Mathew:2017drw, Mathew:2018nvo, Mathew:2020wnx}. As discussed earlier, LQGUP effects only become more apparent with $\alpha_0 > 10^{17}$, since any value below that results to deviations that are too small to distinguish from numerical uncertainty. Therefore, we investigate the implied effects of LQGUP by using the aforementioned values of $\alpha_0$, with the emphasis that such values, though not necessarily within the bound $\alpha_0 < 10^{17}$, are still within the other suggested bounds, and from which we can more noticeably see the behavioral effects of GUP.
	Thus, equations (\ref{Eq8})-(\ref{Eq10}) can be rewritten as:
	\begin{equation}\label{Eq12}
	n=\frac{ \pi m_e^3 c^3}{(2\pi \hbar)^3}\int_0^{\xi} 8\frac{\tilde p^2 d \tilde p}{(1-\sigma \tilde p)^4} = \frac{K}{m_e c^2} \tilde N
	\end{equation}
	\begin{equation}\label{Eq13}
	\epsilon_e=\frac{ \pi m_e^4 c^5}{(2\pi \hbar)^3}\int_0^{\xi}8\tilde{E}\frac{\tilde p^2 d \tilde p}{(1-\sigma \tilde p)^4} = K \tilde \epsilon_e
	\end{equation}
	\begin{equation}\label{Eq14}
	P=\frac{ \pi m_e^4 c^5}{(2\pi \hbar)^3}\int_0^{\xi} 8(\tilde{E}_F-\tilde E)\frac{\tilde p^2 d \tilde p}{(1-\sigma \tilde p)^4} = K \tilde P,
	\end{equation}
	where $\widetilde{E}=\sqrt{1+{\widetilde{p}}^2}$, ${\widetilde{E}}_F=\sqrt{1+\xi^2}$, and $K = \frac{ \pi m_e^4 c^5}{(2\pi \hbar)^3}$ (note that $K$ has units of pressure). $\tilde N$, $\tilde \epsilon_e$, and $\tilde P$ refers to the dimensionless number density, dimensionless electron energy density, and dimensionless pressure respectively. Evaluating equations (\ref{Eq12}) and (\ref{Eq14}) yields the following dimensionless quantities:
	\begin{equation}\label{Eq15}
	\tilde N(\xi)=\frac{8 \xi^3}{3(1-\xi \sigma)^3}
	\end{equation}
    \begin{equation}\label{Eq16}
    \begin{split}
    \widetilde{P}\left(\xi\right)=&\frac{1}{3\sigma^4}4\Bigg\{-2\sqrt{1+\xi^2}\sigma-\frac{2\sqrt{1+\xi^2}\sigma}{\left(-1+\xi\sigma\right)^3}-\frac{6\sqrt{1+\xi^2}\sigma}{\left(-1+\xi\sigma\right)^2}-\frac{6\sqrt{1+\xi^2}\sigma}{-1+\xi\sigma}\\
    &+\frac{\sigma\left(6+11\sigma^2+2\sigma^4\right)}{\left(1+\sigma^2\right)^2}\\
    &+\frac{\left(\sqrt{1+\xi^2}\sigma\left(6+11\sigma^2+2\sigma^4-3\xi\sigma\left(5+9\sigma^2+2\sigma^4\right)\right)\right)}{\left(\left(-1+\xi\sigma\right)^3\left(1+\sigma^2\right)^2\right)}\\
    &+\frac{\left(\sqrt{1+\xi^2}\sigma\left(\xi^2\sigma^2\left(11+20\sigma^2+6\sigma^4\right)\right)\right)}{\left(\left(-1+\xi\sigma\right)^3\left(1+\sigma^2\right)^2\right)}\\
    &-6  \sinh^{-1} \left[\xi\right]-\frac{3\left(2+5\sigma^2+4\sigma^4\right)\ln\left[1-\xi\sigma\right]}{\left(1+\sigma^2\right)^\frac{5}{2}}\\
    &-\frac{3\left(2+5\sigma^2+4\sigma^4\right)\ln\left[\sigma+\sqrt{1+\sigma^2}\right]}{\left(1+\sigma^2\right)^\frac{5}{2}}\\	&+\frac{3\left(2+5\sigma^2+4\sigma^4\right)\ln\left[\xi+\sigma+\sqrt{1+\xi^2}\sqrt{1+\sigma^2}\right]}{\left(1+\sigma^2\right)^\frac{5}{2}} \Bigg\}.
    \end{split}
    \end{equation}
	The total energy density $\epsilon$  is then obtained from the equation:
	\begin{equation}\label{Eq17}
	\epsilon=\rho c^2+\epsilon_e
	\end{equation}
	where the mass density $\rho$ is equal to:
	\begin{equation}\label{Eq18}
	\rho(\xi)=m_u	 \mu_e n = \frac{K}{qc^2} \tilde{N}(\xi)
	\end{equation}
	where $q=\frac{m_e}{(\mu_e m_u)}$, $m_u = 1.6605 \times 10^{-24} \text{g}$, which is the atomic mass, and $\mu_e = A/Z=2$, where $A$ is the mass number and $Z$ is the atomic number (assuming a He white dwarf configuration). The total energy density is:
	\begin{equation}\label{Eq19}
	\epsilon=\frac{K}{q} \tilde{N}(\xi)+K \tilde \epsilon_e = \frac{K}{q} \tilde{\varepsilon},
	\end{equation}
	where $\widetilde{\varepsilon}$ is the dimensionless total energy density and is given by:
	\begin{equation}\label{Eq20}
	\tilde{\varepsilon}= \tilde{N}(\xi) + q \tilde \epsilon_e
	\end{equation}
	\begin{figure}[htb!]
		\includegraphics[width = \textwidth]{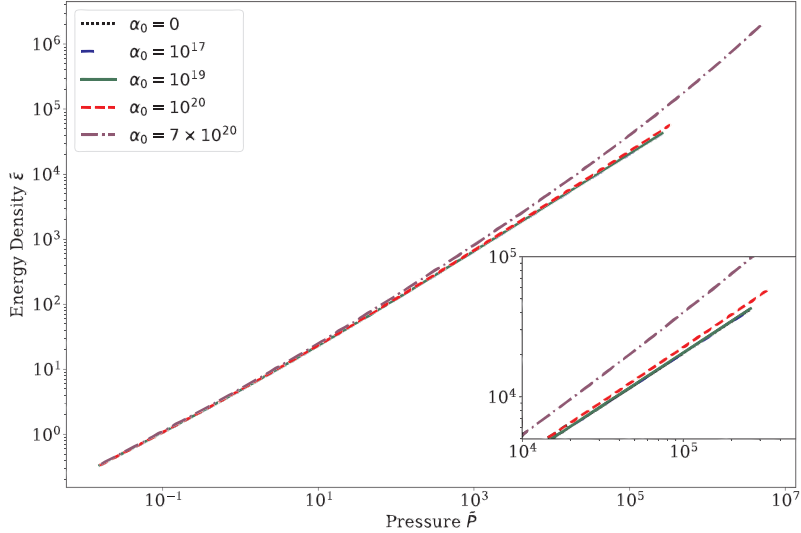}
		\centering
		
		\caption{We show the equation of state of linear-quadratic GUP-modified white dwarfs for $\alpha_0=0$, $10^{17}$, $10^{19}$, $10^{20}$, $7 \times 10^{20}$. \label{Fig2.1}}
	\end{figure}
	It should be noted that the LQGUP-modified equations (\ref{Eq12}), (\ref{Eq13}), and (\ref{Eq14}) are singular at $\tilde p = \frac{1}{\sigma}$. This implies that there is a constraint with regards the momentum values such that $0 \le \tilde p <\frac{1}{\sigma}$. This maximum $\tilde p$ is of the same order of magnitude as the maximum momentum shown in \cite{Abac:2020drc,Nozari:2012gd}. Since $\xi$ is the Fermi momentum for a white dwarf, the LQGUP structure equation for the white dwarf can be safely calculated in the said range, which is sufficient in obtaining the mass-radius relation of white dwarfs. Moreover, it has recently been discussed in Ref. \cite{Prasetyo:2022uaa}, the first law of thermodynamics may be violated when GUP is incorporated in the equation of state, and is only preserved when considering anisotropic pressure. From the first law of thermodynamics, the following should be satisfied: $\mu = (\epsilon+P)/\rho$,
    where $\mu = d\epsilon/d\rho$ is the total chemical potential. With the GUP correction of the order $(1-\sigma \tilde p)^{-4}$, we obtain $\sigma << 1$ even for the largest GUP parameter considered, which is $7 \times 10^{20}$ ($\sigma \sim 0.029$). This means that the GUP corrections considered have negligible impact in the numerical computations since the first law of thermodynamics can only be significantly violated for $\sigma>>1$, which is not the case in this study. 
    
    Equations (\ref{Eq16}) and (\ref{Eq20}) will be used to generate the plot for the equation of state for the LQGUP-modified white dwarfs. The equation of state for the zero-temperature ideal electron gases for various values of the LQGUP parameter $\alpha_0$ is shown in Figure \ref{Fig2.1}. From the plots of the equations of states, it can be seen that the plot for $\alpha_0 = 10^{17}$ and $\alpha_0 = 0$ are practically the same, which means that the effect of linear quadratic GUP for a GUP parameter of $10^{17}$ is negligible. It can also be seen from the plot that there is a slight increase in the slope for $\alpha_0=10^{19}$, which can be seen clearly for larger values of the pressure $\tilde{P}$. A more notable increase of the slope is visible for both $\alpha_0 = 10^{20}$ and $\alpha_0 = 7 \times 10^{20}$ for large values of $\tilde{P}$, which means that the effect of linear-quadratic GUP is significant for $\alpha_0 = 10^{19}$ and $\alpha_0 = 10^{20}$, and is much greater for $\alpha_0 = 7 \times 10^{20}$. From the plots, we can say that the EOS softens with the presence of LQGUP. 
	
	\section{White Dwarf Structure}
	\label{section:White Dwarf Structure}
	In this section, we use the results from Chapter 2 to modify the TOV equations and obtain the mass-radius relations of white dwarfs with LQGUP.
	\subsection{Tolman-Oppenheimer-Volkoff (TOV) Equations}
	\label{Tolman-Oppenheimer-Volkoff (TOV) Equations}
	The structure of white dwarfs can be approximated by the Newtonian equations of stellar structure. However, for the most massive white dwarfs, which are affected by GUP, general relativity (GR) has to be taken into account \cite{Schutz:2003nr}. From the Einstein field equations, and using the metric $ds^2 = e^{\nu}c^2 dt^2 - e^{\mu}dr^2 -r^2 (d\theta^2+\sin^2 \theta d \phi^2)$, one can obtain the structure of GR known as the TOV equations \cite{Tolman:1939jz,Oppenheimer:1939ne}. It has been suggested that a modification to the TOV equations may arise from the modification of the effective metric due to GUP. However, this modification to the TOV equations is outside the scope of the study since we are only interested on the phenomenological effects of LQGUP and mass-radius relations from the standard TOV equations, which is consistent to the approaches of Ref. \cite{Abac:2020drc,Abac:2021jpf,Mathew:2018nvo, Mathew:2020wnx}. The TOV equations are given by \cite{Abac:2020drc,Mathew:2020wnx,Tolman:1939jz,Oppenheimer:1939ne}:
	\begin{equation}\label{2.26}
	\frac{dP}{dr} = -\frac{G}{c^2 r}(\epsilon + P)\frac{m + \frac{4 \pi r^3}{c^2}}{r-\frac{2GM}{c^2}}, \quad P(r=0) = P_c, \quad P(r=R_{\star}) = 0,
	\end{equation}
	\begin{equation}\label{2.27}
	\frac{dm}{dr} = \frac{4 \pi}{c^2} \epsilon r^2, \quad m(r=0) = 0, \quad m(r=R_{\star}) = M_{\star},
	\end{equation}
	where (\ref{2.26}) is the statement of hydrostatic equilibrium, while (\ref{2.27}) is the mass continuity equation together with their boundary conditions. $R_{\star}$ and the $M_{\star}$ are the star radius and star mass, respectively. From the boundary conditions, we can see that the pressure is greatest $P_c$ at the center of the star and gradually decreases to zero at $r=R_{\star}$. On the other hand, the mass at the center of the star is zero, and as we move away from the center of the star, the star accumulates mass until it reaches a mass equal to $M_\star$ at $r=R_{\star}$.
	
	We can then re-express the TOV equations (\ref{2.26}) and (\ref{2.27}) in terms of the dimensionless mass given by $v\equiv\frac{m}{m_0}$, dimensionless radius $\eta\equiv\frac{r}{r_0}$, where $m_0 \equiv \frac{q^2c^4}{G\sqrt{4\pi G\ K}} $ and $r_0 \equiv \frac{qc^2}{\sqrt{4\pi G\ K}}$. With this, we start with
 
	\begin{equation}\label{2.28}
	\frac{d\widetilde{P}}{d\eta}=\frac{d\widetilde{P}}{d\xi}\frac{d\xi}{d\eta} \rightarrow \frac{d\xi}{d\eta}\ =\frac{1}{\frac{d\widetilde{P}}{d\xi}}\frac{d\widetilde{P}}{d\eta}\ ,
	\end{equation}
	where $\frac{d\widetilde{P}}{d\xi}$ and $\frac{d\widetilde{P}}{d\eta}$ are:
	\begin{equation}\label{2.29}
	\frac{d\widetilde{P}}{d\xi}=\frac{\xi}{\sqrt{1+\xi^2}}\widetilde{N}\left(\xi\right)
	\end{equation}
	and
	\begin{equation}\label{2.30}
	\frac{d\widetilde{P}}{d\eta}=\frac{-\left(\widetilde{\varepsilon}+q\widetilde{P}\right)}{\eta}\frac{v+q\widetilde{P}\eta^3}{\eta-2qv}	
	\end{equation}
	respectively. We can then obtain the dimensionless hydrostatic equilibrium equation for $\frac{d\xi}{d\eta}$
	\begin{equation}\label{2.31}
	\frac{d\xi}{d\eta} =-\frac{\sqrt{1+\xi^2}}{\xi} \frac{1+q\sqrt{1+\xi^2}}{\eta} \frac{v+q \tilde P \eta^3}{\eta - 2qv}, \quad \xi(\eta=0) = \xi_c, \quad \xi(\eta = \eta_R) = 0.
	\end{equation}
	Equation (\ref{2.27}) can then be rewritten in terms of dimensionless quantities and is given by:
	\begin{equation}\label{2.32}
	\frac{m_0dv}{r_0d\eta}=\frac{4\pi}{c^2}\frac{K}{q}\widetilde{\varepsilon}\left(\eta r_0\right)^2,
	\end{equation}
	which can be simplified into the dimensionless mass-continuity equation:
	\begin{equation}\label{2.33}
	\frac{dv}{d\eta}=\frac{4\pi}{c^2}\frac{K}{q\ m_0}\widetilde{\varepsilon}{r_0}^3\eta^2=\frac{4\pi}{c^2}\frac{K}{q\ m_0}\widetilde{\varepsilon}\eta^2\frac{\left(\frac{qc^2}{\sqrt{4\pi G\ K}}\right)^3}{\frac{q^2c^4}{G\sqrt{4\pi G\ K}}},
	\end{equation}
	and we finally obtain:
	\begin{equation}\label{2.34}
	\frac{dv}{d\eta} = \tilde \varepsilon\eta^2, \quad v(\eta=0) = 0, \quad v(\eta = \eta_R) = v_R
	\end{equation}
	where $v_R$ and $\eta_R$ are the dimensionless final star mass ($v_R = M_{\star}/m_0$) and final star radius ($\eta_R = R_{\star}/r_0$), respectively. Notice that in this case, we have redefined our TOV equations in terms of the Fermi momentum $\xi$ instead of the usual pressure $p$, for convenience in dealing with the dynamical instability analysis later on. This is different from approach of Ref. \cite{Abac:2020drc} (where the pressure TOV equation was not reexpressed in terms of the Fermi momentum), but should nevertheless yield the same results.
	\subsection{Mass-Radius Relations}
	\label{Mass-Radius Relations}
	
	\begin{figure}[htb!]
		\includegraphics[width = \textwidth]{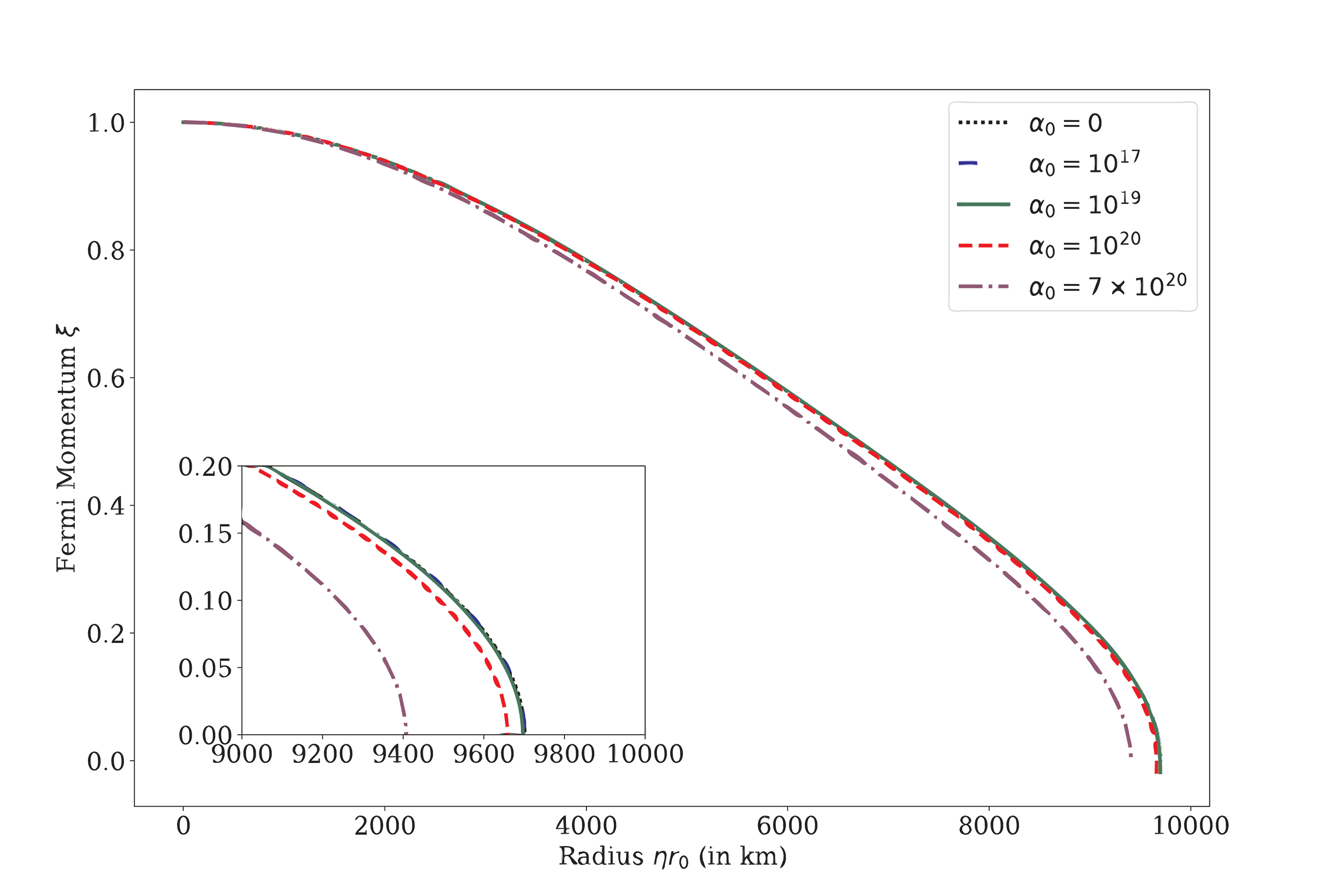}
		\centering
		\caption{The figure shows the dimensionless Fermi momentum $\xi$ as a function of radius $\eta \, r_0$ (in km) for GUP parameter values $\alpha_0=0$, $10^{17}$, $10^{19}$, $10^{20}$, $7 \times 10^{20}$, given an initial central momentum $\xi_c = 1.0$.\label{Fig2.2} The final star radius $\eta_R$ is obtained when $\xi = 0$.}
	\end{figure}
 
	Figure \ref{Fig2.2} shows the Fermi momentum versus $\eta$ plot for both the LQGUP-modified white dwarfs and the white dwarf with no linear-quadratic GUP corrections. The figure shows that the Fermi momentum of the star decreases from the center of the white dwarf and vanishes at the radius of the white dwarf. We also see that the GUP $\alpha_0 = 10^{17}$ appears to have a negligible effect on the white dwarf. On the other hand, the white dwarfs with $\alpha_0 =$ $10^{19}$, $10^{20}$, and $7 \times 10^{20}$ have variations to their plots with respect to the ideal case, where they have smaller final star radius than the white dwarfs with no LQGUP effects. This trend is more evident when looking at the GUP parameter of $7 \times 10^{20}$, where the plot is shown to have a smaller value of $\eta_R$.
	\begin{figure}[htb!]
		\includegraphics[width = \textwidth]{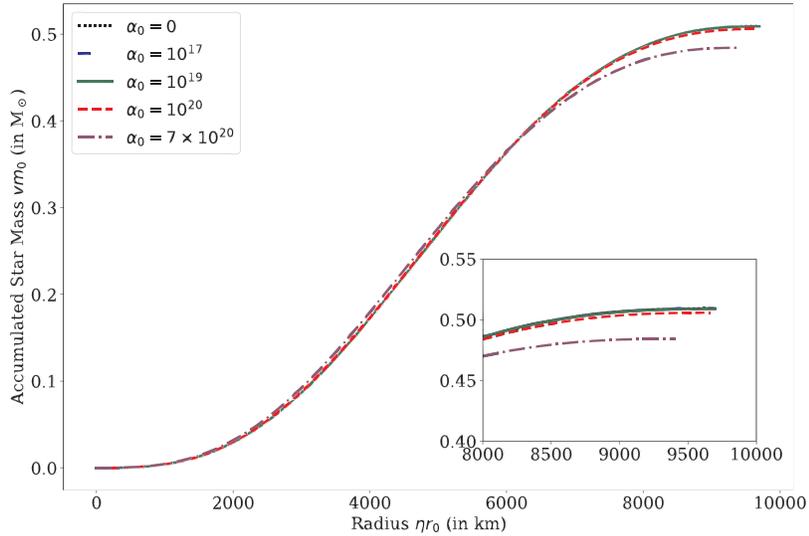}
		\centering
		\caption{The accumulated star mass $v m_0$ (in $M_\odot$) as a function of accumulated star radius $\eta \, r_0$ (in km) for GUP parameter values $\alpha_0=0$, $10^{17}$, $10^{19}$, $10^{20}$, $7 \times 10^{20}$. \label{Fig2.3}, with inital $\xi_c = 1.0$. Note that in this case, the star accumulates mass until it reaches its final star mass $v_R$ at $\eta = \eta_R$. The value of $v_R$ gets smaller with increasing $\alpha_0$.}
	\end{figure}
 
	Figure \ref{Fig2.3} shows the $v$ versus $\eta$ plot for varying values of the GUP parameter $\alpha_0$. The plot with $\alpha_0 = 10^{17}$ shows negligible effects as it shows a similar trend to the plot for $\alpha_0 = 0$. On the other hand, $\alpha_0=10^{19}$, and $\alpha_0=10^{20}$ show significant LQGUP effects on the white dwarfs, where LQGUP appears to decrease the star mass of the white dwarf, which is notable at larger values of $\eta$. The plot also shows that for $\alpha_0 = 7 \times 10^{20}$, $v_R$ is much smaller compared to those with smaller $\alpha_0$.
	\begin{figure}[htb!]
		\includegraphics[width = \textwidth]{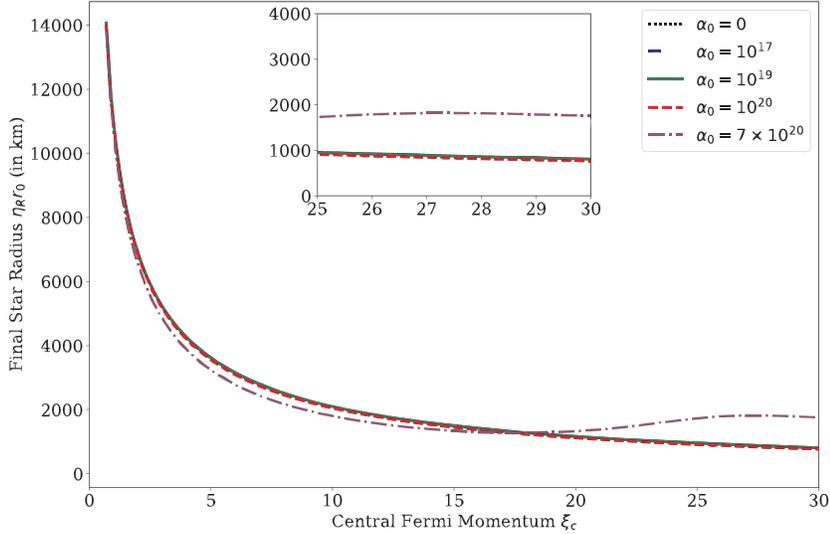}
		\centering
		\caption{Final star radius $\eta_R \, r_0$ (in km) as a function of the central Fermi momentum $\xi_c$ for GUP parameter values $\alpha_0=0$, $10^{17}$, $10^{19}$, $10^{20}$, $7 \times 10^{20}$. \label{Fig2.4}}
	\end{figure}
	We then solve the TOV equations for a chosen range of values of the central Fermi momentum $\xi_c$, taking caution that $\xi_c < 1/\sigma$, and plot the final radius $\eta_R$ and star mass $v_R$ as a function of $\xi_c$. The star radius $\eta_R$ versus central Fermi momentum plot $\xi_c$ (shown in Figure \ref{Fig2.4}) indicates that various values of $\xi_c$ correspond to different star radius values. Larger values of $\xi_c$ corresponds to a smaller star radius, while smaller values of $\xi_c$ corresponds to a more prominent star radius. This checks out with the fact that a larger $\xi_c$ indicates a more compact object, thus, a smaller radius. It should be noted that for a GUP parameter of $7\times 10^{20}$, $\eta_R$ decreases faster at low values of $\xi_c$ compared with smaller $\alpha_0$. However, $\eta_R$ reaches a minimum at some $\xi_c$ before increasing for larger values of $\xi_c$, while for the other $\alpha_0$, $\eta_R$ just decreases in value in this range of central Fermi momentum $\xi_c \in [0,25]$. Note that the range of the central Fermi momenta that we used is well within the constraints we placed for the Fermi momentum. This is sufficient to describe the trend of the mass radius relation, as well as the dynamical instability analysis for LQGUP-modified white dwarfs. 
	\begin{figure}[htb!]
		\includegraphics[width = \textwidth]{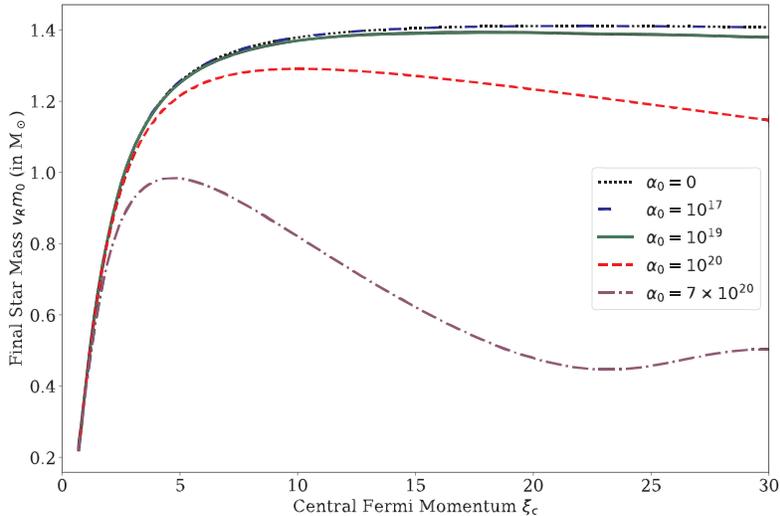}
		\centering
		\caption{Final star mass $v_R \, m_0$ (in $M_\odot$) as a function of dimensionless central Fermi momentum $\xi_c$ for GUP parameter values $\alpha_0=0$, $10^{17}$, $10^{19}$, $10^{20}$, $7 \times 10^{20}$. \label{Fig2.5}}
	\end{figure}
 
	Figure \ref{Fig2.5} corresponds to the star mass versus the central Fermi momentum plot, which shows the noticeable effects of GUP on the star mass. The plot shows that the effect of GUP is to decrease the maximumam star mass, that is the Chandrasekhar limit. The GUP parameter $\alpha_0 = 10^{17}$ exhibits negligible effects to the maximum star mass, while $\alpha_0 = 10^{19}$ shows a noticeable decrease and a drastic decrease of the maximum star mass for $\alpha_0 = 10^{20}$. For $\alpha_0=7 \times 10^{20}$, $v_R$ drastically decreases smaller values of $\xi_c$, however, for larger values of Fermi momentum $\xi_c$, $v_R$ is shown to increase.
	\begin{figure}[htb!]
		\includegraphics[width = \textwidth]{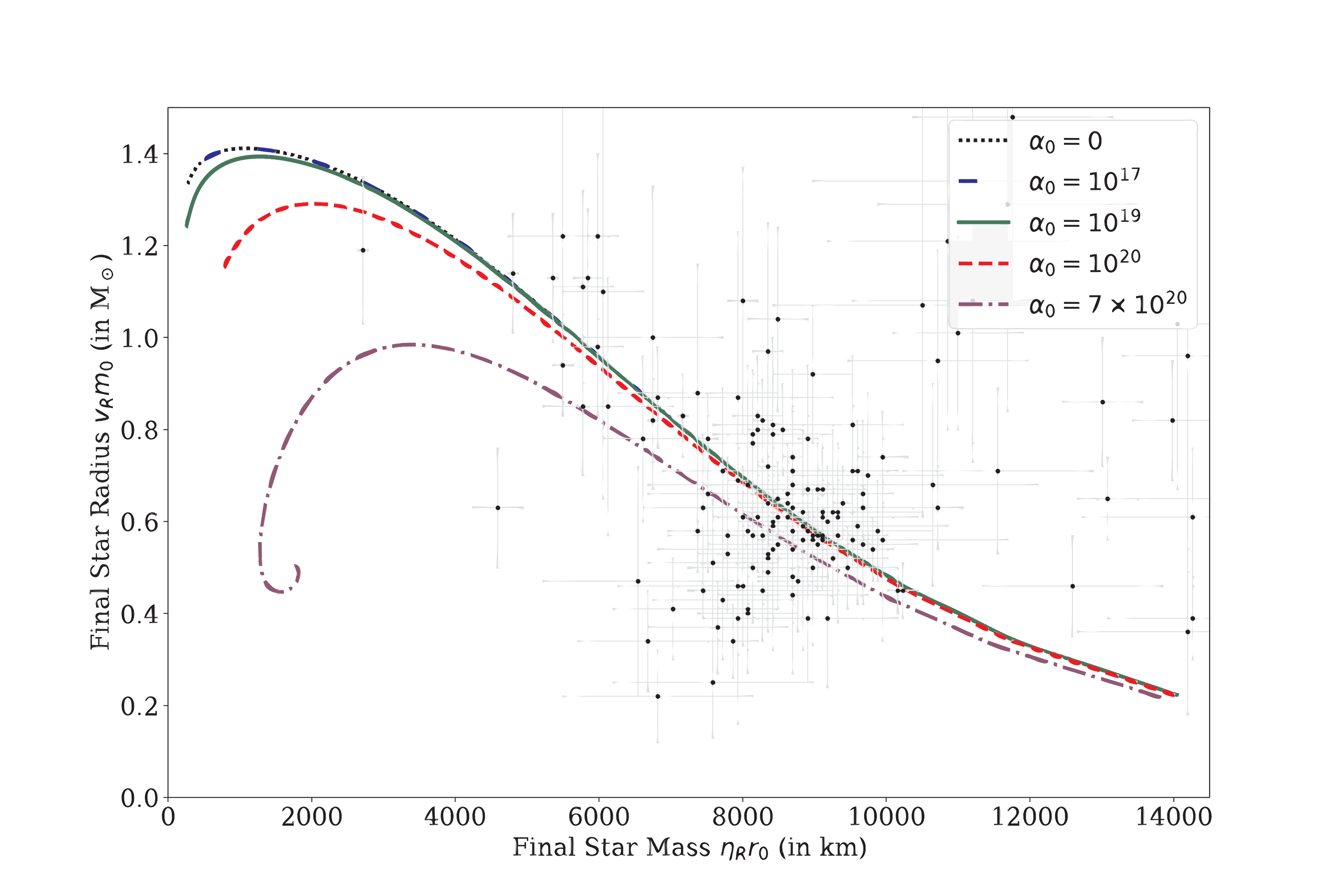}
		\centering
		\caption{The plot shows the final star mass $v_R \, m_0$ (in $M_\odot$) as a function of final star radius $\eta_R \, r_0$ (in km) for GUP parameter values $\alpha_0=0$, $10^{17}$, $10^{19}$, $10^{20}$, $7 \times 10^{20}$ together with observational data taken from Ref. \cite{B_dard_2017}. \label{Fig2.6}}
	\end{figure}
 
    	The mass-radius relations can be seen in Figure \ref{Fig2.6}, (together with observational data taken from Ref. \cite{B_dard_2017}) which contains the parametric plots between star mass vs star radius. The plots show that as the star radius decreases, the star mass increases, which means that the white dwarf is heavier when it's dimensionally smaller, demonstrating its compact nature. Examining the plot from the largest star radius decreasing towards zero (right to left), we see that the star mass eventually reaches a maximum value, before decreasing towards zero radius.  The plots also show that for larger values of the GUP parameter $\alpha_0$, the maximum mass is smaller, while $\alpha_0 = 10^{17}$ shows that the effects of the GUP parameter are negligible. There is also a larger limiting radius, which is the radius by which the star reaches its maximum mass, for the larger GUP parameters. In addition, for a GUP parameter of $7 \times 10^{20}$, as the final radius of the white dwarf becomes smaller, the mass increases but eventually decreases after reaching the maximum mass of the white dwarf. At the lower values of the final radius, the plot starts to spiral, where both the final mass and final radius increases. It should be noted that direct comparisons of the observation data and our results to determine the most compatible GUP parameter is not straightforward to establish and one has to take caution in doing so, mainly because the regime where GUP is expected to dominate have very little observational data. As another note of caution, we also state that the GUP parameters used are larger than some of the bounds discussed earlier (particularly $\alpha_0 < 10^{17}$), and do not necessarily mean that such GUP parameters exist, since the main point of this study is to see how the changes from the ideal case would look like given an increasing (non-zero) value of the LQGUP parameter; in other words, the effects may be exaggerated than what one may expect in nature. From Figure \ref{Fig2.6}, it can be seen that most of the observation data can be found from the low to intermediate mass stars and have very few data points for high mass stars, where the GUP effects is very noticeable. In addition, the assumption of ideal electron gas for the equation of state of the white dwarfs makes it difficult to constrain the possible values of the GUP parameter. Nevertheless, we perform a chi-square analysis in the vein of Ref. \cite{Belfaqih:2021jvu} to see which GUP parameter looks ``most compatible" with current observational data. From Ref. \cite{Belfaqih:2021jvu}, we can use chi-square analysis to investigate whether the observation data used in this study favors the LQGUP model. The chi-square between the observational data and theoretical data is given by: $ \Delta \chi_i^2= (M-M_i)^2/\sigma^2_{M,i} + (R-R_i)^2/\sigma^2_{R,i}$, where $M_i$ and $R_i$ are the observation mass and radius data, $M$ and $R$ are from our LQGUP-modified mass-radius relations, while $\sigma_{M,i}$ and $\sigma_{r,i}$ are the mass and radius standard deviation of the observational data, respectively. From this, we extract the minimum $\Delta \chi^2_i$ value for each GUP parameter value, and calculate the $\chi^2$ value by $\chi^2=\Sigma_i \Delta X_i^2$. We find the following chi-square values $\chi^2 = [432.610, 432.606, 432.1, 428.7, 558.0]$ for $\alpha_0 = [0, 10^{17}, 10^{19}, 10^{20}, 7\times 10^{20}]$. The $\chi^2$ values decrease from $\alpha_0=0$ to $10^{20}$ before jumping to a large value at $7 \times 10^{20}$. More sophisticated equations of state, such as the inclusion of other compositions beside the ideal electron gas, as well as improved independent mass and radius measurements (where neither relies on an existing mass-radius relation to calculate the radius from the mass or vice-versa) is needed to further constrain the GUP parameters, as discussed on Ref. \cite{Danarianto:2023rff}. The increasing sophistication of the EoS to make it more realistic, not to mention the incorporation of GUP (both in the EoS and in other cases, the metric itself of the underlying theory of gravity), are non-trivial, and is outside the scope of this study. 
     
	\begin{table}[]
		\centering
		\begin{tabular}{|l|l|l|}
			\hline
			$\alpha_0$      & $v_R$ ($M\odot$) & $\eta_R$(km)     \\ \hline
			0               & 1.411 & 1077.129 \\ \hline
			$10^{17}$         & 1.411  & 1077.129 \\ \hline
			$10^{19}$         & 1.393 & 1269.338 \\ \hline
			$10^{20}$         & 1.289 & 2033.731 \\ \hline
			$7\times 10^{20}$ & 0.984 & 3409.005 \\ \hline
		\end{tabular}
		\caption{Values for the maximum value of star mass $v_R$ with the corresponding star radius $\eta_R$ for different values of the GUP parameter.}
		\label{table1}
	\end{table}
	The table shows the value of maximum mass and minimum radius of the white dwarf for every $\alpha_0$. We see that for $\alpha_0 = 10^{17}$ are similar with $\alpha_0 = 0$, which means that the effect of LQGUP is negligible. In addition, we can see that the maximum value of $v_R$ decreases with increasing value of the GUP parameter. On the other hand, the corresponding $\eta_R$ increases when $\alpha_0$ is larger. This means that for larger values of the GUP parameter $\alpha_0$, LQGUP-modified white dwarfs have a lower mass limit. The behaviours observed in the structure equations and mass radius relations are similar to those obtained from Ref. \cite{Abac:2020drc}, albeit using a different approach, where we wrote the TOV equations in terms of the Fermi momentum $\xi$. We will see in the next section that white dwarfs beyond the maximum mass limit are unstable configurations. 
	
	\section{Dynamical Instability Analysis}
	\label{section: Dynamical Instability Analysis}
	In Subsection 4.1, we review some general results in analyzing the dynamical instability of white dwarfs, following the procedures laid out by Ref. \cite{Chandrasekhar:1931ih,Chandrasekhar:1934in,Chandrasekhar:1964zza,Mathew:2017drw,Mathew:2018nvo,Mathew:2020wnx,1966ApJ...145..505B}. In Subsection 4.2, we present our own derived expressions for the dynamical instability analysis containing LQGUP modifications.
	\subsection{Radial Oscillations and Dynamical Instability of White Dwarfs}
	\label{Radial Oscillations and Dynamical Instability of White Dwarfs}
	For the dynamical stability analysis, we start with the metric interior of the star, which can be expressed as \cite{Chandrasekhar:1964zza,Mathew:2017drw,Mathew:2020wnx}:
	\begin{equation}\label{3.1}
	ds^2 = e^{\nu+\delta \nu}c^2 dt^2 - e^{\mu + \delta \mu}dr^2 -r^2 (d\theta^2+\sin^2 \theta d \phi^2).
	\end{equation}
	From the equation, $\nu\left(r\right)$ and $\mu\left(r\right)$ here refers to the equilibrium metric potentials, while $\delta\nu\left(r,t\right)$ and $\delta\mu\left(r,t\right)$ refers to the perturbations due to small radial Lagrangian displacements $\zeta\left(r,t\right)$ given by $\zeta\left(r,t\right)=r^{-2}e^{\frac{\nu}{2}\psi\left(r\right)e^{i\omega t}}$ which induces perturbations $\delta P\left(r,t\right)$ and $\delta\epsilon\left(r,t\right)$. The radial oscillation equation can then be obtained in the Sturm-Liouville form and is given by\cite{Mathew:2020wnx,1966ApJ...145..505B}:
	\begin{equation}\label{3.2}
	\frac{d}{dr} \Big(U \frac{d\psi}{dr}\Big) +(V+\frac{\omega^2}{c^2}W)\phi = 0,
	\end{equation}
	With the boundary conditions $\psi=0$ at $r=0$, and $\delta P=-e^\frac{\nu}{2}\left(\frac{\gamma P}{r^2}\right)\frac{d\psi}{dr}=0\,\, \text{at}\,\, r=R$, the terms in equation (3.2) are:
	\begin{equation}\label{3.3}
	U(r) = e^{\frac{(\mu + 3 \nu)}{2}}\frac{ \gamma P}{r^2},
	\end{equation}
	\begin{equation}\label{3.4}
	V(r) = -4\frac{e^\frac{\mu+3\nu}{2}}{r^3}\frac{dP}{dr}-\frac{8\pi G}{c^4}\frac{e^{3\frac{\mu+\nu}{2}}}{r^2}P\left(P+\epsilon\right)+\frac{e^\frac{\mu+3\nu}{2}}{r^2}\frac{1}{P+\epsilon}\left(\frac{dP}{dr}\right)^2,
	\end{equation}
	\begin{equation}\label{3.5}
	W\left(r\right)=\frac{e^\frac{3\mu+\nu}{2}}{r^2}\left(P+\epsilon\right),
	\end{equation}
	where $\gamma$ is the adiabatic index equal to:
	\begin{equation}\label{3.6}
	\gamma=\frac{\epsilon+P}{P}\left(\frac{dP}{d\epsilon}\right).
	\end{equation}
	
	Distributing $\psi$ to equation (\ref{3.2}) and then integrating, where we let $ \psi' = \frac{d \psi}{dr}$, we obtain:
	\begin{equation}\label{3.10}
	J\left[\psi\right]=\int_{0}^{R}{U\psi^{\prime2}-V\psi^2-\frac{\omega^2}{c^2}W\psi^2}dr.
	\end{equation}
	Minimizing equation (\ref{3.7}) with respect to $\psi$, we can then obtain the Sturm-Liouville equation in equation (\ref{3.2}), which means we have a variational basis for determining the lowest characteristic eigenfrequency and is expressed as:
	\begin{equation}\label{3.11}
	\frac{\omega_0^2}{c^2}=\min_{\psi\left(r\right)}\frac{\int_{0}^{R}{(U}\psi^{\prime2}-V\psi^2)dr}{\int_{0}^{R}W\psi^2dr}
	\end{equation}
	
	From equation (\ref{3.12}), we will then obtain a set of $\omega_0^2$ values corresponding to a range of central Fermi momentum $\xi$. A star with an $\omega_0^2$ that is greater than zero means that the star is stable, while a star that has a negative $\omega_0^2$ value means that dynamical instability is present for that certain central Fermi momentum value. The trial function $\psi$ of the fundamental mode can be approximated in the form $\psi(r)=c_0r^3$ \cite{Chandrasekhar:1964zza,Mathew:2020wnx,1968ApL.....2..253W} The equation for the central mass density can be found in equation (\ref{Eq18}). With this, the mass density at which the white dwarf shows gravitational collapse will be the critical density $\rho_c$* and can be identified with the zero eigenfrequency solution of the equation.
	
	The interior Schwarzschild metric potentials, which satisfy Einstein's field equations, are given by \cite{Mathew:2020wnx,Tolman:1939jz,Oppenheimer:1939ne}
	\begin{equation}\label{4.9}
	e^{-\mu(r)} =1-\frac{2GM}{c^2r}	,
	\end{equation}
	and
	\begin{equation}\label{4.10}
	e^{\nu}=\left(1-\frac{2GM}{c^2R}\right)\left(\exp{\left[-2\int_{0}^{P\left(r\right)}\frac{dP}{\varepsilon+P}\right]}\right).
	\end{equation}
	
	\subsection{Dynamical Instability Analysis for LQGUP-modified White Dwarfs}
	\label{Dynamical Instability Analysis for LQGUP-modified White Dwarfs}
	We then rewrite the expressions in Section 4.1, this time taking into account the dimensionless quantities we defined in Section \ref{section: Linear quadratic GUP-modified White Dwarfs} together with LQGUP modification. The dimensionless adiabatic index $\tilde \gamma$ can be obtained using the definition of $\frac{dP}{d\tilde \varepsilon}$ given by:
	\begin{equation}\label{3.7}
	\frac{dP}{d\tilde \varepsilon} =\left[\frac{\xi^2\left(1-\xi\sigma\right)}{3\sqrt{1-\xi^2}\ \left(1+q\sqrt{1+\xi^2}\right)}\right],
	\end{equation}
	where $\tilde \gamma$ will be:
	\begin{equation}\label{3.8}
	\tilde \gamma=\frac{\frac{K}{q}\widetilde{\varepsilon}+K\widetilde{P}}{K\widetilde{P}}\left(\frac{d \tilde P}{d\tilde \varepsilon}\right)=\frac{\widetilde{\varepsilon}+q\widetilde{P}}{\widetilde{P}}\left[\frac{\xi^2\left(1-\xi\sigma\right)}{3\sqrt{1-\xi^2}\ \left(1+q\sqrt{1+\xi^2}\right)}\right],
	\end{equation}
	and can be simplified to:
	\begin{equation}\label{3.9}
	\tilde \gamma=\tilde{N} \xi^2\frac{\left(1-\sigma\ \xi\right)}{3\tilde P  \sqrt{1+\xi^2}}
	\end{equation}
	
	Equation (\ref{3.11}) can then be rewritten in terms of dimensionless quantities given by \cite{Mathew:2020wnx}:
	\begin{equation}\label{3.12}
	\omega_0^2=\left(\frac{qc^2}{r_0^2}\right)\frac{\tilde{U}+\tilde{V}}{\tilde{W}},
	\end{equation}
	where the equations for $\tilde{U}$, $\tilde{V}$, and $\tilde{W}$, are given by:
	\begin{equation}\label{3.13}
	\tilde{U}=\int_{0}^{\eta_R}{e^{\left(\mu+3\nu\right)/2}\frac{\gamma\widetilde{P}}{\eta^2}}\psi^{\prime2}d\eta,
	\end{equation}
	\begin{equation}\label{3.14}
	\tilde{V}=\int_{0}^{\eta_R}{\frac{e^{\left(\mu+3\nu\right)/2}}{\eta^2}\left[\frac{4}{\eta}\frac{d\widetilde{P}}{d\eta}+2qe^\mu\widetilde{P}\left(\widetilde{\varepsilon}+q\widetilde{P}\right)-\frac{q}{\widetilde{\varepsilon}+q\widetilde{P}}\left(\frac{d\widetilde{P}}{d\eta}\right)^2\right]\psi^{2}}d\eta,
	\end{equation}\
	\begin{equation}\label{3.15}
	\tilde{W}=\int_{0}^{\eta_R}{e^{\left(3\mu+\nu\right)/2}\frac{\widetilde{\varepsilon}+q\widetilde{P}}{\eta^2}}\psi^{2}d\eta,
	\end{equation}
	respectively. The interior Schwarzschild metric potential from Equations (\ref{4.9}) and (\ref{4.10}) can also be re-expressed in terms of dimensional quantities given by:
	\begin{equation}
	e^{-\mu}=1-\frac{2Gvm_0}{c^2\eta r_0}=1-\frac{2Gv\frac{\left(qc^2\right)^2}{\left(G^{3/2}\sqrt{4\pi K}\right)}}{c^2\eta\frac{qc^2}{\sqrt{4\pi G\ K}}}=1-2q\frac{v}{\eta}
	\end{equation}
	and
	\begin{equation}
	e^{\nu}=\left(1-2q\frac{v_R}{\eta_R}\right)\left(exp{\left[-2\int_{0}^{P\left(r\right)}\frac{qKd\widetilde{P}}{K(\widetilde{\varepsilon}+q\widetilde{P})}\right]}\right),
	\end{equation}
	which can be simplified into
	\begin{equation}
	e^{\nu\left(\eta\right)}=\left(1-2q \frac{v_R}{\eta_R}\right)\left[\frac{\left(1+q\right)^2}{\left(1+q\sqrt{1+\xi^2}\right)^2}\right].
	\end{equation}
	\begin{figure}[htb!]
		\includegraphics[width = \textwidth]{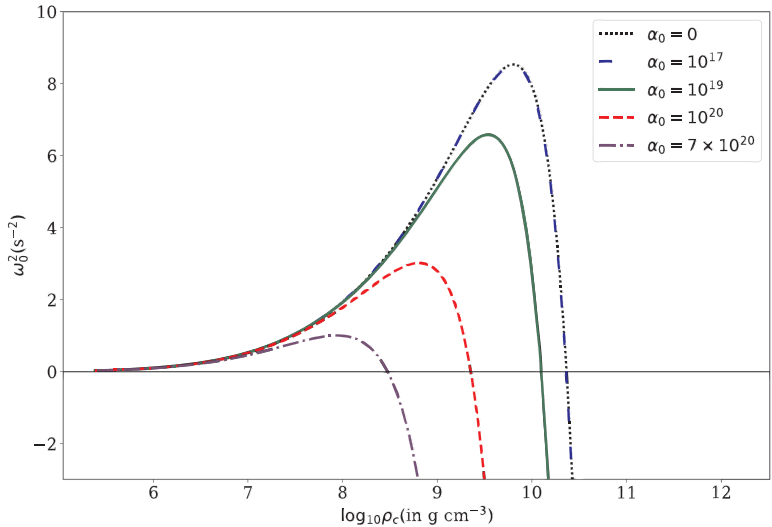}
		\centering
		\caption{The figure above shows the eigenfrequency of the fundamental mode $\omega_0^2$ versus central mass density $\rho_c$ for for GUP parameter values $\alpha_0=0$, $10^{17}$, $10^{19}$, $10^{20}$, $7 \times 10^{20}$. \label{Fig4.6}}
	\end{figure}
 
	Plotting the numerical integrations from the equations above gives us the eigenfrequency $\omega_0^2$ versus central density  $\rho_c$ plot for varying values of the GUP parameter $\alpha_0$ shown in Figure \ref{Fig4.6}. From the plot, we can see that white dwarfs with central densities $\rho_c \lesssim {10}^7 \hspace{.1cm} \text{g} \hspace{.1cm} \text{cm}^{-3}$ have overlapping plots, which means that the effect of the GUP parameter does not have a notable impact on the white dwarfs in the indicated range of central densities. This also means that all the values of the GUP parameter in the range of $0\le\alpha_0\le{10}^{17}$ will yield similar results, which are overlapping plots with the ideal case ($\alpha_0=0$) and have approximately the same value of the central density by which gravitational collapse begins. This is also apparent in the mass radius relation where low mass white dwarfs are not significantly affected by GUP. The plots of the different GUP parameters start to deviate in the higher density regime, where a larger GUP parameter such as for $\alpha_0={10}^{20}$, $\alpha_0={10}^{19}$, and $\alpha_0={7 \times 10}^{20}$ deviates at a lower $\rho_c$ than the other values of $\alpha_0$. The plot for $\alpha_0 = 7 \times 10^{20}$ deviates from the ideal case at an even smaller $\rho_c$, reaching the maximum value of $\omega_0^2 \approx 1$  and is shown to decrease immediately from there at smaller values of the central density. On the other hand, a GUP parameter equal to ${10}^{17}$ has negligible effects on the white dwarfs. A larger GUP parameter shows that the white dwarf has a smaller central density by which its eigenfrequency reaches a maximum value. This is true for GUP parameters ${10}^{19}$, $10^{20}$, and ${7 \times 10}^{20}$, while the plot of $\alpha_0 = {10}^{17}$ has a large $\omega_0^2$ value, and corresponding $\rho_c$* value. However, it should be noted that all values of the GUP parameter yield similar behaviors, which show a descending value for eigenfrequency in higher central densities. This means that there exists a zero eigenfrequency solution at critical central densities $\rho_c$* for different values of the LQGUP parameter (even at very large $\alpha_0$), beyond which, at $\omega_0^2 \le 0$ the white dwarfs will undergo gravitational collapse. This is very much in contrast with the results from Ref. \cite{Mathew:2020wnx}, which found a range of values of the quadratic GUP parameter for which the white dwarf can still undergo instability, and beyond which, the white dwarf is super-stable and can have any arbitrary mass, which is very much in contrast with astrophysical observations. This is also means that LQGUP removes any need for quadratic GUP to impose additional constraints in the super-stable case just so that the white dwarf can retain its maximum limiting mass even in a super-stable configuration. Additionally, we can infer from the mass-radius relations or the dynamical instability analysis that even with LQGUP in GR, the condition \cite{Glendenning:1997wn} $(dM/d\rho_c) < 0$ is preserved, unlike the one with QGUP but also in the context of GR. It may then be interesting to see how LQGUP could affect this condition for beyond-GR theories, but this is outside the scope of the study and we reserve delving deeper into this question for future extensions.
 
	\begin{table}[]
		\centering
		\begin{tabular}{|l|l|l|l|}
			\hline
			$\alpha_0$ & $\rho_c$* & $v_R$ ($M\odot$) & $\eta_R$(km) \\ \hline
			0          &     1.036 $\times 10^{11}$      &      1.415            &      1025.493        \\ \hline
			$10^{17}$    &  1.036 $\times 10^{11}$         &        1.415          &        1033.486      \\ \hline
			$10^{19}$    &   1.009 $\times 10^{11}$       &        1.396          &       1243.393       \\ \hline
			$10^{20}$    &   9.332 $\times 10^{10}$       &        1.291          &       2031.892       \\ \hline
			$7 \times 10^{20}$    &   8.433 $\times 10^{10}$      &        0.985          &       3407.039       \\ \hline
		\end{tabular}
		\caption{Values for the critical mass density $\rho_c$*, star mass $v_R$, star radius $\eta_R$ for varying values of the GUP parameter $\alpha_0$ at the onset of dynamical instability.}
		\label{table2}
	\end{table}
 
	Table \ref{table2} shows the values of the critical mass density $\rho_c$*, star mass, star radius for various values of $\alpha_0$. The table shows that the critical mass density values, star radius, and star mass are similar for $\alpha_0 = 0$ and $\alpha_0 = 10^{17}$. However, for GUP parameter $\alpha_0 = 10^{19}$, $\alpha_0 = 10^{20}$, and $\alpha_0 = 7 \times 10^{20}$, the critical mass density and final mass are shown to decrease with increasing values of the GUP parameter. This means that larger values of the GUP parameter reduces the final mass before it undergoes gravitational collapses. On the other hand, the larger the GUP parameter, the larger the final radius of the white dwarf before it collapses. Since the final mass of the white dwarf decreases with increasing values of the GUP parameter, the Chandrasekhar mass limit of $1.42 \text{M}_\odot$ remains unaltered (for $\alpha_0 \le 10^{17}$). In addition, the maximum value of the critical central density from the table is much smaller than the nuclear matter density  of around $10^{14} \hspace{.1cm} \text{g} \hspace{.1cm} \text{cm}^{-3}$, thus the assumption of a free fermionic equation of state holds true to the stable regime of the white dwarfs. Comparing the table above to Table \ref{table1}, we can see that the corresponding $v_R$ and $\eta_R$ values of the peaks from the mass-radius relation plot in Figure \ref{Fig2.6} are practically similar to Table \ref{table2} for the same values of $\alpha_0$. This means that from the mass-radis relation plots, we can identify that instability occurs for larger $\alpha_0$ values at larger star radius and smaller star mass. With this, we know that dynamical instability happens after reaching the maximum mass of the star, when examining the plots from the largest star radius decreasing towards zero. When the radius of the star is smaller than the corresponding radius of the star after reaching its maximum mass, dynamical instability occurs. 
	\begin{figure}[htb!]
		\includegraphics[width = \textwidth]{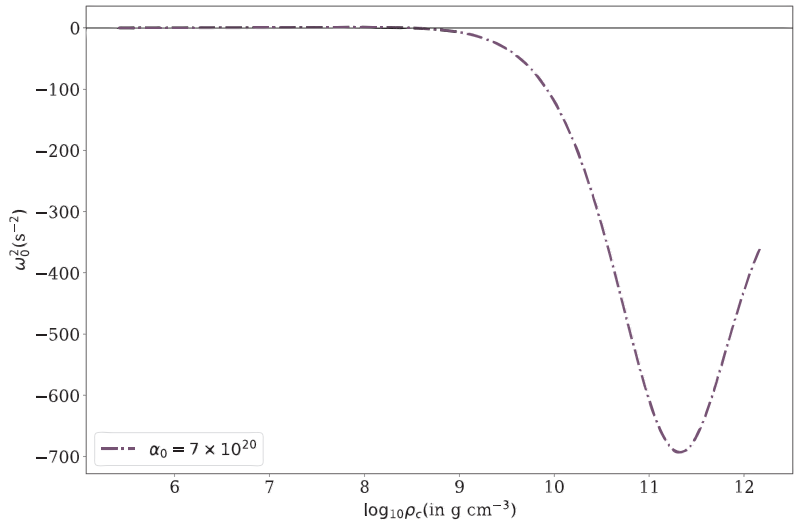}
		\centering
		\caption{Eigenfrequency of the fundamental mode $\omega_0^2$ versus central mass density $\rho_c$ for different values of $\alpha_0$ \label{Fig4.7}}
	\end{figure}
	Figure 8 shows the complete plot of $\omega_0^2$ versus central mass density $\rho_c$  for $\alpha_0 = 7 \times 10^{20}$, which shows all the values of $\omega_0^2$ below zero. From this plot, we can see that $\omega_0^2$ decreases from zero, until it reaches the minimum value of $\omega_0^2$, where it increases again from there. This behavior confirms the spiraling that occurs in the mass-radius plot, which are still deemed unstable. \\
 	\begin{figure}[htb!]
		\includegraphics[width = \textwidth]{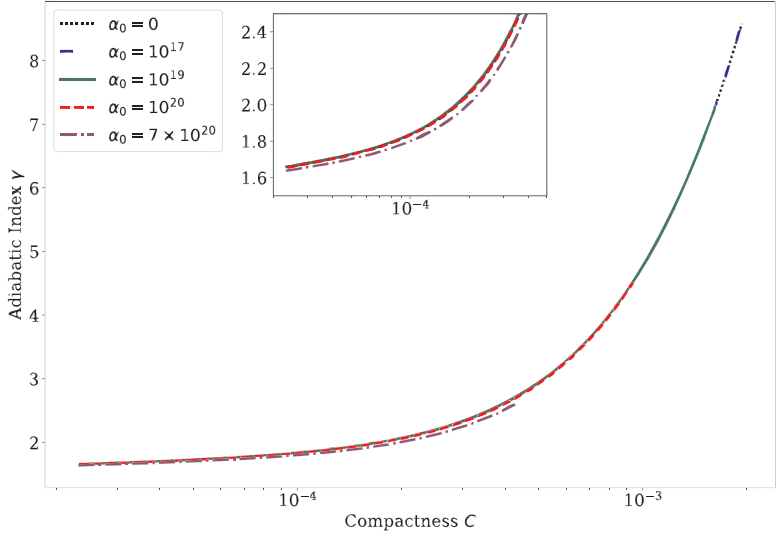}
		\centering
		\caption{Plot for adiabatic index $\gamma$ as a function of compactness $C$ for different values of $\alpha_0$ \label{Fig4.8}}
	\end{figure}
 
    In addition, we solve for the adiabatic index for different values of $\alpha_0$, and plot in \ref{Fig4.8} adiabatic index $\gamma$ as a function of the compactness $C$, where $C=\frac{G M}{R c^2}$, plotted up to the compactness of the maximal stable mass in the mass-radius relations. We observe very similar trends for the adiabatic index for different (relatively) smaller values of $\alpha_0$, whereas at $\alpha_0 = 7 \times 10^{20}$, there is significant deviation of $\gamma$ from $\alpha_0 = 0$ case at relatively large compactness. This also helps confirm our observation in Table \ref{table2} where the star gets less compact with relatively larger values of $\alpha_0$. In investigating the effects of LQGUP on the adiabatic index of the system, we first note that we have imposed a constraint to our Fermi momenta $\xi$, i.e. $0\le \xi<1/\sigma$ (where $\sigma = m_e \alpha_0/M_p < 1$) to prevent the onset of singularity, and as already noted above, this is already sufficient in obtaining the mass-radius relations of the white dwarfs (as well as dynamical instability analysis) from a desired range of $\xi_c$. In the limiting case then, we take the upper bound $\xi \rightarrow 1/\sigma$, instead of the usual $\xi \rightarrow \infty$. This yields $\lim \gamma|_{\xi\rightarrow 1/\sigma} = 2/3$, implying that this limiting value is independent of the LQGUP parameter $\sigma$. Note that for $\sigma \rightarrow 0$ (no GUP), $\lim \gamma|_{\xi\rightarrow \infty} = 4/3$. This result is also different from the result of Ref. \cite{Mathew:2020wnx}, where they obtained a result of $\lim \gamma|_{\xi \rightarrow \infty} = (\pi/16)\beta^3$, which is dependent on the value of the quadratic GUP parameter $\beta$.
    
	\section{Conclusions and Recommendations}
	\label{chapter: Conclusions and Recommendations}

	The mass-radius relations plots from Ref. \cite{Abac:2020drc} are successfully simulated in this study, albeit using a different approach. In general, the LQGUP decreases the mass of the white dwarf for a given radius, and this is particularly most manifested in the most massive white dwarfs. The results of the mass-radius relations and dynamical instability analysis show that dynamical instability sets in at different values of $\eta_R$ and $v_R$ for various values of the GUP parameter $\alpha_0$. The mass-radius relations plot shows us that after reaching a certain maximum mass, the white dwarfs will become unstable and will continuously decrease its mass and radius due to instability. On the other hand, based on the results from dynamical instability analysis, we find that the dynamical instability sets in for all values of the GUP parameters $\alpha_0$, albeit at different mass limits. Therefore, unlike in the case of QGUP \cite{Mathew:2020wnx}, there seems to be no bound to the LQGUP parameter for which dynamical instability does not set in; in other words the condition \cite{Glendenning:1997wn} $(dM/d\rho_c) < 0$ is satisfied for the wide range of LQGUP parameters used here. This verifies the results obtained from the mass-radius relations plots. In addition, we find that the Chandrasekhar mass limit is preserved for LQGUP-modified white dwarfs as seen from the onset of gravitational collapse, where as we increase the value of $\alpha_0$, the mass limit will tend to decrease. The results show that when the eigenfrequency of the fundamental mode $\omega_0^2$ is plotted against the central density $\rho_c$ for various values of the GUP parameter $\alpha_0$, the eigenfrequencies $\omega_0^2$ continuously decrease up to the critical central density $\rho_c$* indicating gravitational collapse. 
 
	It can be observed that the eigenfrequency for a GUP parameter $\alpha_0={10}^{17}$ coincides with the plot for the ideal case, which has no GUP correction. With this, for GUP parameters within the range of $\alpha_0=0$ and $\alpha_0={10}^{17}$, the critical central density $\rho_c*$ for the onset of gravitational collapse will be approximately the same. For values of the GUP parameter in the range $0\le\alpha_0\le{10}^{17}$, their corresponding mass-radius relation plot shows maximal points corresponding to the Chandrasekhar mass limit at around $1.42 \text{M}_\odot$. On the other hand, larger values of $\alpha_0$ is shown to have a lower maximum value when compared to $\alpha_0$ values less than $10^{17}$, which implies that the mass limit by which instability sets in is lower. It is given that upon reaching a mass beyond the Chandrasekhar mass limit, the star would collapse and will form a more compact object. With this, the LQGUP-modification supports gravitational collapse by decreasing the Chandrasekhar mass limit of the white dwarf. This is opposite to the results obtained when using QGUP \cite{Mathew:2020wnx}, where at some values of the GUP parameter $\beta_0$, the white dwarf will remain stable at masses beyond the Chandrasekhar mass limit. We also note that in both the LQGUP case studied here and QGUP case, the dynamical instability treatment is within the context of general relativity (GR). Therefore, incorporating GUP itself onto GR via an effective metric or to other modified gravity theories would be an interesting extension to this study to see whether or not $(dM/d\rho_c) < 0$ still holds in any of these cases \cite{Belfaqih:2021jvu,Ong:2023jkp, Maulana:2019sgd}.
	
	The nature of the GUP parameter and phase space volume approach used affects the resulting equation of state and mass radius relations of white dwarfs. Thus, Ref. \cite{Ong:2023jkp} notes that it is important to work with the full expression of the GUP correction from the LQGUP model instead of using the series truncation, which is one of the limitations for this study. With this, future studies can work on the effect the GUP parameter to the white dwarf structures using higher-order GUP approaches as suggested from Ref. \cite{Chung:2019raj,Gregoris:2022ekv}. In addition, modifications to the TOV equations due to the effect of GUP to effective metric is expected. Thus, analyzing the effects of the modifications to the TOV equation for GUP-modified white dwarfs is recommended for future studies. From Ref. \cite{Prasetyo:2022uaa}, it should be noted that there is a possibility for GUP modifications to the equation of state could violate the first law of thermodynamics for large values of the GUP parameter. One solution for this is to assume anisotropic pressures for the system, which wasn't implemented for this study and should be taken account for future work. Furthermore, the study considers an ideal electron gas, which is one of the limitations of the study. Hence, considering refinements and modifications on the equation of state including many-body interactions, envelopes, finite temperatures, and magnetic fields is needed for a more realistic description of the white dwarf's interior \cite{Danarianto:2023rff,Saltas:2018mxc}, and with the addition of GUP, this will pose a challenge for next studies, not to mention the inclusion also of GUP onto theories of gravity from which the structures of the white dwarfs will be solved. Aside from that, one can also consider denser compact objects such as neutron stars and black holes which may increase the chance of constraining the GUP parameters, since these objects warp spacetime more significantly than white dwarfs \cite{Abac:2021jpf}.  

    Interestingly, we also find that the qualitative effect of LQGUP onto the mass-radius relations is similar to that of dark matter, where it has been shown in Ref. \cite{Abac:2021txj,Panotopoulos:2017idn,Das:2018frc} that the mass of the neutron star is decreased with the inclusion of dark matter. It would be worthwhile to investigate how dark matter will affect such system that is considered in this study, and how to break apart the potential degeneracies that can occur if the effects of both dark matter and LQGUP were to be considered.
    
    The works of Ref. \cite{Ong:2018zqn} suggested that the quadratic GUP parameter $\beta$ can be negative, where the negative GUP parameter can prevent the white dwarf from having arbitrary large masses. This negative value has also been suggested in the uncertainty relations in the crystal lattice \cite{Jizba:2009qf}, as well as the Magueijo-Smolin formulation \cite{Magueijo:2001cr} of doubly special relativity (DSR) \cite{Scardigli:2019pme}. On the other hand, a negative GUP parameter $\alpha_0$ have been considered by Ref. \cite{Belfaqih:2021jvu}, but in the context of entropic gravity from Ref. \cite{Verlinde:2010hp}, which is not considered in this study. Additionally, dynamical instability analysis can be used for other astronomical systems such as neutron stars. Dynamical instability analysis can be used to analyze the radial oscillations of neutron stars and compare its behavior to other compact objects when modified by LQGUP. Lastly, future studies can explore using other numerical methods for solving the equations above to improve the accuracy and efficiency of the calculations. With this, using larger values of $\alpha_0$ would be possible, which will be helpful in identifying the upper bound of the LQGUP parameter to be used for white dwarfs.

	\section*{Acknowledgments}
	The authors thank the Department of Science and Technology, Philippines and Ironwood Corporation for providing financial support for this project.

\bibliography{References.bib}














\end{document}